\documentclass{aa} 
\usepackage{times,graphics,psfig}
\begin{document}
\sloppy

\overfullrule5pt
\thesaurus{07(08.01.3;   % Stars: atmospheres
              08.05.1;   % Stars: early-type
              08.13.2;  % Stars: mass-loss  
              11.04.1;   % Galaxies: distances and redshifts
              12.04.3)}   % Cosmology: distance scale
   \title{The Wind Momentum - Luminosity Relationship of Galactic
          A- and B-Supergiants}

   \author{R.P. Kudritzki \inst{1,2,3}
         \and J.Puls \inst{1}
         \and D.J. Lennon \inst{1,4}
         \and K.A. Venn \inst{5}
         \and J. Reetz \inst{1}
	 \and F. Najarro \inst{9}
         \and J.K. McCarthy \inst{6} 
         \and A. Herrero \inst{7,8}}

   \offprints{R.P. Kudritzki}

   \institute{ Institut f\"ur Astronomie und Astrophysik der Universit\"at
              M\"unchen, Scheinerstr.1, D-81679 M\"unchen, Germany
        \and Max-Planck-Institut f\"ur Astrophysik, Karl-Schwarzschild-Str. 1,
             D-85740, Garching, Germany
        \and University of Arizona, Steward Observatory, 933 N. Cherry Av.,
             Tucson AZ 85721, USA
        \and Isaac Newton Group of Telescopes, Apartado de Correos 368,
             E-38700 Santa Cruz de La Palma, Spain
        \and Department of Physics and Astronomy, Macalaster College,
             St.Paul, MN 55101, USA
        \and Palomar Observatory, California Institute of Technology,
             Pasadena, CA 91125, USA
        \and Instituto de Astrof\'{\i}sica de Canarias,
             c/V\'{\i}a Lactea s/n, E-38200 La Laguna, Tenerife, Spain
        \and Departamento de Astrofisica, Universidad de La Laguna, 
             Avda. Astrof. Francisco Sanchez s/n, E-38071 La Laguna, Spain
        \and Instituto de Estructura de la Materia, CSIC, Serrano 121, 28006, 
	     Madrid, Spain}

   \date{Received ; accepted  }
   \titlerunning{WLR of A- and B-Supergiants}
   \authorrunning{R.P. Kudritzki {\it et al.}}

   \maketitle

   \begin{abstract}

The Balmer lines of four A Ia - supergiants (spectral type A0 to A3) and
fourteen B Ia and Ib - supergiants (spectral type B0 to B3) in the solar
neighbourhood are analyzed by means of NLTE unified model atmospheres to
determine the properties of their stellar winds, in particular their wind
momenta. As in previous work for O-stars (Puls et al. \cite{pul96}) a
tight relationship between stellar wind momentum and luminosity (``WLR'') is
found. However, the WLR varies as function of spectral type. Wind momenta
are strongest for O-supergiants, then decrease from early B (B0 and B1) to
mid B (B1.5 to B3) spectral types and become stronger again for
A-supergiants. The slope of the WLR appears to be steeper for A- and mid
B-supergiants than for O-supergiants. The spectral type dependence is
interpreted as an effect of ionization changing the effective number and the
line strength distribution function of spectral lines absorbing photon
momentum around the stellar flux maximum. This interpretation needs to be
confirmed by theoretical calculations for radiation driven winds.

The ``Pistol-Star'' in the Galactic Centre, an extreme mid B-hypergiant
recently identified as one of the most luminous stars (Figer et al.
\cite{fig99}) is found to coincide with the extrapolation of the mid
B-supergiant WLR towards higher luminosities. However, the wind momentum of
the Luminous Blue Variable P Cygni, a mid B-supergiant with extremely strong
mass-loss, is 1.2 dex higher than the WLR of the ``normal'' supergiants.
This significant difference is explained in terms of the well-known stellar
wind bi-stability of supergiants very close to the Eddinton-limit in this
particular range of effective temperatures. A-supergiants in M31 observed
with HIRES at the Keck telescope have wind momenta compatible with their
galactic counterparts.

The potential of the WLR as a new, independent extragalactic distance
indicator is discussed. It is concluded that with ten to twenty objects,
photometry with HST and medium resolution spectroscopy with 8m-telescopes
from the ground distance moduli can be obtained with an accuracy of about
$0\fm1$ out to the Virgo and Fornax clusters of galaxies.


\end{abstract}

\section{Introduction}

The theory of radiation driven winds (Castor et al.~\cite{cak75},
Pauldrach et al.~\cite{pau86}, Kudritzki et al.
\cite{kud89}) predicts a tight relationship between the total mechanical
momentum flow $\dot{M}v_{\infty}$ contained in the stellar wind outflow and
the luminosity $L$ of the mass-losing star

\begin{equation}
 \dot{M}v_{\infty} \propto R_{\ast}^{-1/2} L^{1/\alpha_{\rm eff}}.
 \label{wlreq}
\end{equation}

\noindent
In Eq.\,\ref{wlreq} $\dot{M}$ is the mass-loss rate, $v_{\infty}$ the
terminal velocity of the stellar wind, and $R_{\ast}$ the stellar radius.
$\alpha_{\rm eff}$ is a dimensionless number of the order of $2/3$ and
represents the power law exponent of the distribution function of line
strengths of the thousands of spectral lines driving the wind. (For a
simplified derivation of Eq.\,\ref{wlreq}, see Kudritzki \cite{kud98}, or
Kudritzki \cite{kud99}).

Analyzing H$_{\alpha}$ line profiles of O-stars in the Galaxy and in the
Magellanic Clouds Puls et al. (\cite{pul96}) were able to demonstrate
that such a relationship does indeed exist for the most luminous O-stars
ranging over 1 dex in stellar luminosity. In addition, Kudritzki {\it et
al.} (\cite{kud97}) found that Central Stars of Planetary Nebulae being
equally hot or hotter than O-stars but two orders of magnitudes less
luminous show wind momenta corresponding to the relationship for O-stars
extrapolated towards lower luminosities.

The existence of such a relationship introduces a variety of new and
interesting aspects. First of all, if reliably calibrated, it can be used to
determine distances by purely spectroscopic means with mass-loss rates and
terminal velocities obtained directly from line profiles yielding absolute
stellar luminosities. Second, together with the well known proportionality
of v$_{\infty}$ to the escape velocity v$_{\rm esc}$ from the stellar
surface (Abbott \cite{abb82}) it provides a simple way to predict stellar
wind energy and momentum input into the interstellar medium from clusters
and associations of hot stars. Third, it may be used to estimate mass-loss
rates, wind energies and momentum along sequences of stellar evolutionary
tracks (for a more detailed discussion of these aspects, see Kudritzki
\cite{kud99}).

With this in mind, we have started to investigate whether the winds of
luminous blue supergiants of spectral types later than O also follow a
relationship as described by Eq.\,\ref{wlreq}. From the viewpoint of theory
this is clearly to be expected, although it is very likely that the
proportionality constant and the exponent $\alpha_{\rm eff}$ will be
different, since the winds of A- and B-supergiants are driven by lines of
different ionization stages. However, our hope is that a similarily tight
relationship may exist. This would allow to extend the simple description of
stellar wind strengths towards later stages of stellar evolution. Even more
important, it would render the possibility to use the optically brightest
``normal'' stars, super- and hypergiants of spectral type A and B, as new
primary extragalactic distances indicators for galaxies far beyond the Local
Group. In a first pragmatic step, Kudritzki et al. (\cite{kud95}) used
mass-loss rates obtained by Barlow \& Cohen (\cite{bar77}) from
IR-photometry and first estimates of terminal velocities to find indeed an
indication for such a relationship at later spectral types.

In this paper, we present the first results of our spectroscopic attempt to
establish the Wind momentum -- Luminosity Relationship (WLR) of galactic
supergiants of spectral type A and B. We concentrate on a small number of
objects with reasonably well determined distances. Optical spectra are used
to obtain the stellar parameters of $T_{\rm eff}, \log g, R_{\ast}, L$.
Then, H$_{\alpha}$ - line profiles are analyzed in detail to determine the
stellar wind properties, in particular the mass-loss rates. Together with
terminal velocities measured from from UV-resonance lines and H$_{\alpha}$
(for A-supergiants) this yields stellar wind momenta as function of the
stellar luminosities.

\section{Model atmospheres for the stellar wind analysis} 

The determination of stellar wind parameters from the Balmer lines of blue
supergiants requires the sophisticated effort of using NLTE unified model
atmospheres with stellar winds and spherical extension, as originally
introduced by Gabler et al. (\cite{gab89}). These new types of model
atmospheres provide a smooth transition from the subsonic, quasi-hydrostatic
photosphere to the supersonic stellar wind and allow to deal simultaneously
with wind contaminated absorption lines like H$_{\gamma}$ and pronounced
stellar wind emission lines such as H$_{\alpha}$. They are also essential
for stars with weak winds, when H$_{\alpha}$ turns into an almost
photospheric absorption line, a situation impossible to deal with
quantitatively by treating photosphere and wind as separated entities. For
our work, we use the new unified model code developed recently by
Santolaya-Rey et al.~(\cite{sph97}, hereafter SPH). This new code
is extremely fast and produces a unified model in few minutes on a work
station, which is crucial for a project aiming at the analysis of many stars
with different stellar parameters and different wind properties. In
addition, it has the advantage of being numerically stable over the entire
domain of O-, B- and A-stars in the HRD, independent of the adopted stellar
wind strengths. The code can deal with extreme emission lines produced by
very strong winds as well as with an entire absorption spectrum for
extremely weak winds. In these latter cases it re-produces perfectly the
line profile results of plane-parallel, hydrostatic NLTE model atmosphere
codes. As SPH have shown, this is only possible because the Sobolev
approximation is avoided and the NLTE multilevel radiative transfer is
treated in the comoving frame of the stellar wind outflow velocity field. In
addition, incoherent electron scattering can be included in the final formal
integral to calculate the line profiles. This is important for low gravity
A-supergiants, where {\it photospheric} electron-scattering produces shallow
and wide emission wings (see SPH and McCarthy et al. \cite{mcc97}).

\begin{table}
  \caption[]{Adopted distance moduli}
  \label{tab:clusters}
  \begin{center}
    \begin{tabular}{l r l}
      \hline
assoc. & m-M & source \\
      \hline

 Per OB1 & 11$\fm$8 & Garmany \& Stencel, 1992 \\
 Ori OB1 &  8$\fm$5 & Blaha \& Humphreys, 1989 \\
 Gem OB1 & 10$\fm$9 & Blaha \& Humphreys, 1989 \\
 Cyg OB1 & 10$\fm$5 & Garmany \& Stencel, 1992 \\
 Cas OB5 & 11$\fm$5 & Garmany \& Stencel, 1992 \\
 Cas OB14& 10$\fm$2 & Blaha \& Humphreys, 1989 \\
 Car OB1 & 12$\fm$5 & see text, section 6 \\ 
 Col 121 &  9$\fm$2 & Blaha \& Humphreys, 1989 \\
 Cep OB2 &  9$\fm$9 & Garmany \& Stencel, 1992 \\

\end{tabular}
\end{center}
\end{table}

A unified model atmosphere is defined by the effective temperature $T_{\rm
eff}$, the gravity $\log g$ and the stellar radius $R_{\ast}$ at the Rosseland
optical depth $\tau_{\rm ross}=2/3$ together with $\dot{M}$, $v_{\infty}$
and the parameter $\beta$ which describes the radial slope of the velocity
field via

\begin{equation}
 v(r) = v_{\infty}(1-b/r)^{\beta}.
\end{equation}

\noindent
The constant $b$ is chosen to guarantee a smooth transition into the
hydrostatic stratification at a prespecified outflow velocity smaller than
the isothermal sound speed (normally 0.1 $v_{\rm sound}$). Observed velocity
fields are usually well represented by this parametrization.

As for the analysis of O-star stellar wind lines in the UV (see Haser {\it
et al.} \cite{has98} for a most recent reference) the assumption of a local
Gaussian microturbulent velocity $v_{\rm t}$ is needed to fit the wind
affected Balmer line profiles of B- and A-supergiants. For simplicity, we
adopt a constant value of $v_{\rm t}$ through the entire atmosphere. We are
aware of the fact that a stratified turbulent velocity field increasing
outwards (as for O-stars, see Haser et al. \cite{has98}) would
probably be more appropriate. However, since the determination of wind
momenta is only little affected by any assumption concerning the
microturbulence, we postpone the inclusion of depth dependent
microturbulence to a later study. In some cases we find that the fit of
H$_{\gamma}$ requires a smaller value of $v_{\rm t}$ than H$_{\alpha}$. We
attribute this to a stratification of the microturbulent velocity.

\begin{table*}
  \caption[]{Spectral types and photometric data of B-supergiants}
  \label{tab:bphoto}
  \begin{center}
    \begin{tabular}{l l l c c c c}
      \hline
 star & spec. type & assoc. & m$_{\rm v}$ & B-V & E(B-V) & M$_{\rm v}$  \\
      \hline
  HD  &               &        &   mag   & mag &  mag   &  mag     \\     
      \hline
  37128 & B0Ia    & Ori OB1  & 1.70 & -0.19 & 0.06 & -6.99 \\
  38771 & B0.5Ia  & see text & 2.04 & -0.18 & 0.06 & -4.80 \\
   2905 & BC0.7Ia & Cas OB14 & 4.16 &  0.14 & 0.31 & -7.00 \\
  13854 & B1Iab   & Per OB1  & 6.47 &  0.28 & 0.50 & -6.70 \\
  13841 & B1.5Ib  & Per OB1  & 7.36 &  0.23 & 0.41 & -5.57 \\
 193183 & B1.5Ib  & Cyg OB1  & 7.00 &  0.44 & 0.64 & -5.47 \\
  41117 & B2Ia    & Gem OB1  & 4.63 &  0.27 & 0.42 & -7.54 \\
  14818 & B2Ia    & Per OB1  & 6.25 &  0.30 & 0.48 & -6.87 \\
  14143 & B2Ia    & Per OB1  & 6.64 &  0.50 & 0.65 & -6.95 \\
 206165 & B2Ib    & Cep OB2  & 4.74 &  0.30 & 0.48 & -6.40 \\
  13866 & B2Ib    & Per OB1  & 7.49 &  0.19 & 0.37 & -5.33 \\
  42087 & B2.5Ib  & Gem OB1  & 5.75 &  0.22 & 0.38 & -6.32 \\
  53138 & B3Ia    & Col 121  & 3.01 & -0.11 & 0.06 & -6.37 \\
  14134 & B3Ia    & Per OB1  & 6.57 &  0.48 & 0.69 & -7.13 \\

\end{tabular}
\end{center}
\end{table*}

The spectral lines of B- and A- supergiants are also broadened by stellar
rotation. We determined rotational velocities $v_{\rm rot}$ from weak metal
lines in the optical spectra by a comparison of observed line profiles with
computations assuming an intrinsically narrow (only thermally broadened)
line which was convolved with a rotational and instrumental broadening
function. For the B-supergiants we compared these results with the values
obtained by Howarth et al. (\cite{how97}) from IUE high resolution
spectra. In most cases our $v_{\rm rot}$ values were somewhat smaller
corresponding to a systematically difference of 12 percent. We attribute
this to the different technique applied but note that for the determination
of wind momenta this small difference is not important.

\section{Spectroscopic data}

Blue and red spectra were available for all of our targets. The B-supergiant
data have been obtained by DJL and are described in Lennon et al.
(1992, 1993) and McErlean et al. (\cite{mce99}). For the A-supergiants
(except HD 92207) the spectra were taken by KAV using the Coude-spectrograph
at the 2.1m-telescope at McDonald Observatory, Texas (see Venn 1995a,b for a
more detailed description of the setup used). For HD92207 Echelle-spectra
obtained with the ESO 50cm-telescope at La Silla (see Kaufer et al.
\cite{kauf96}) were provided by A. Kaufer (LSW Heidelberg). For the analysis
presented below it is important to note that the S/N of all spectra used is
better than 100 and that the spectral resolution at H$_{\alpha}$ exceeds the
photospheric rotational velocities.

\section{Distances}

To calibrate the WLR the stellar radii and, therefore, the distances must be
known. For A- and B-supergiants this is a major source of uncertainty, since
most of these objects in the solar neighbourhood are already to remote too
allow a direct measurement of parallaxes. Therefore, distances have to be
determined from association and cluster membership. For our work we have
selected objects belonging to Per OB1, Ori OB1, Gem OB1, Cyg OB1, Cas OB5,
Cas OB14, Car OB1, Cep OB2and Col 121 (see Table\,\ref{tab:clusters}).

One object, HD 38771 ($\kappa$ Ori) appears to be in the foreground of its
association according to the Hipparcos-catalogue, which gives $\pi= 4.52$
and $\sigma=0.77$ (in $\mu$arcsec) for parallax and standard deviation,
respectively. Although also quite uncertain in view of the recent discussion
of Hipparcos data (see Narayanan \& Gould \cite{nara99a};
Reid \cite{reid98}) and hard to believe in view of the average absolute 
magnitude of early B Ia supergiants, we make use of this distance.

\section{Spectral analysis of B-supergiants}

14 B-supergiants of luminosity class I and spectral types B0 to B3 have been
selected from the list of objects studied by Lennon et al.
(\cite{len92}, \cite{len93}) according to the
criteria that reasonable measurements of $v_{\infty}$ from ultraviolet
spectra and estimates of the distance modulus are available. For normal
B-supergiants of spectral types later than B3 it is difficult to determine
$v_{\infty}$ from the UV (see Howarth et al. \cite{how97}). This
explains why we have not included spectral types B4 to B9 in this study,
although H$_{\alpha}$ profiles indicate the significant presence of winds in
many cases.

\begin{table*}
  \caption[]{Stellar parameters and stellar wind properties of B-supergiants}
  \label{tab:stellb}
  \begin{center}
    \begin{tabular}{l c c c c c c c r c c}
      \hline
 star & T$_{\rm eff}$ & R$_{\ast}$ & $\log g$ & $\log L/{\rm L}_{\odot}$ & v$_{\rm rot}$ &
v$_{\rm t}$ & $\beta$ & v$_{\infty}$ & $\dot{M}$ & $\log D_{\rm mom}$ \\
      \hline
  HD  &    kK     & R$_{\odot}$&  cgs  &                   &   km/s    &
 km/s   &         & km/s         &$10^{-6}$ M$_{\odot}$/yr& cgs \\     
      \hline
  37128 & 28.5 & 35.0 & 3.00 & 5.86 & 80 & 20 & 1.25 & 1600. & 2.40 & 29.15 \\
  38771 & 27.5 & 13.0 & 3.00 & 4.94 & 80 & 10 & 1.00 & 1350. & 0.27 & 27.93 \\
   2905 & 24.0 & 41.0 & 2.70 & 5.70 & 60 & 20 & 1.35 & 1100. & 2.30 & 29.01 \\
  13854 & 23.5 & 35.3 & 2.70 & 5.53 & 80 & 20 & 1.50 & 1000. & 0.78 & 28.46 \\
  13841 & 22.0 & 22.9 & 2.70 & 5.04 & 75 & 20 & 3.0  & 1015. & 0.038 & 27.06 \\
 193183 & 22.5 & 21.4 & 2.70 & 5.02 & 70 & 20 & 3.0  &  545. & 0.035 & 26.74 \\
  41117 & 19.5 & 61.7 & 2.25 & 5.70 & 40 & 20 & 1.0  &  500. & 0.85 & 28.32 \\
  14818 & 20.0 & 45.4 & 2.40 & 5.47 & 70 & 40 & 2.0  &  650. & 0.25 & 27.83 \\
  14143 & 20.0 & 47.1 & 2.30 & 5.51 & 65 & 10 & 1.75 &  650. & 0.30 & 27.92 \\
 206165 & 20.0 & 36.5 & 2.50 & 5.28 & 80 & 25 & 2.5  &  700. & 0.060 & 27.20 \\
  13866 & 20.5 & 22.3 & 2.60 & 4.90 & 85 & 15 & 2.0  &  870. & 0.020 & 26.71 \\
  42087 & 20.5 & 35.2 & 2.50 & 5.30 & 60 & 40 & 3.0  &  735. & 0.11 & 27.48 \\
  53138 & 18.5 & 39.6 & 2.30 & 5.22 & 60 & 40 & 2.5  &  620. & 0.095 & 27.37 \\
  14134 & 18.0 & 56.2 & 2.20 & 5.48 & 60 & 20 & 3.0  &  465. & 0.15 & 27.52 \\
\end{tabular}
\end{center}
\end{table*}

The photometric data (from Blaha \& Humphreys \cite{bh89}), association or
cluster membership and spectral type (from the compilation by Howarth {\it
et al.} \cite{how97}, see references therein) are given in
Table\,\ref{tab:bphoto}. To calculate the de-reddened absolute magnitude
M$_{\rm v}$ a value of R$_{\rm v}$=3.1 has been adopted for the ratio of
interstellar extinction A$_{\rm v}$ to reddening E(B-V) for all objects
except those in Per OB1 (R$_{\rm v}$=2.75) and Cep OB2 (R$_{\rm v}$=2.6).
These latter values have been proposed by Cardelli et al.
(\cite{card89}).

\subsection{Stellar parameters}

The stellar parameters are obtained from the quantitative analysis of the
spectrum. We adopt the effective temperatures
determined in the NLTE analysis by McErlean et al. (\cite{mce99})
using unblanketed hydrostatic model atmospheres. Their T$_{\rm eff}$ scale
is based on the HeII/I ionization equilibrium for T$_{\rm eff} \geq$ 26000K
and on the SiIV/III ionization equilibria for the cooler B-supergiants in
our sample. Then, we calculate unified model
atmosphere fluxes for the V-filter to obtain stellar radii from M$_{\rm v}$
as described by Kudritzki (\cite{kud80}). Since unified model atmospheres
are spherically extended, the predicted flux depends in principle also on
the radius adopted for the model atmosphere calculations. This requires an
iteration with regard to the stellar radius. However, as long as the
mass-loss rates are not extremely large, the radius dependence of the model
atmosphere flux is very weak so that the iteration is not neccessary or
converges very quickly, if needed. No attempts are made to re-determine
effective temperatures using unified model atmosphere calculations of
HeI/HeII or SiIII/IV lines. This is possible, in principle, with our present 
version of the
SPH-code but would increase the computational effort sigificantly for the
large number of objects studied. Therefore, we postpone a completely 
self-consistent
study iterating stellar paramters, chemical abundances and mass-loss rates
at this stage.

In their NLTE study using hydrostatic, planeparallel model atmospheres
McErlean et al. (\cite{mce99}) have also determined stellar gravities
$\log g$ from a fit of the observed H$_{\gamma}$ profiles. In our analysis, we
start with their values but, since stellar wind emission as a function of the
mass-loss rate does not only affect H$_{\alpha}$ but also higher Balmer
lines such as H$_{\beta}$ and H$_{\gamma}$ (Gabler et al.
\cite{gab89}, Puls et al. \cite{pul96}, Kudritzki \cite{kud98}), we always checked the fits of these latter two Balmer lines and
modified the gravity accordingly, if necessary. In most of the cases the
$\log g$ values obtained by McErlean et al. (\cite{mce99}) were 
sufficiently accurate.

A normal helium abundance of N(He)/N(H)=0.1 was adopted for the calculation
of all model atmospheres. The analysis of the observed He{\sc i} spectra of
B-supergiants still gives somewhat divergent results depending on the lines
used for the abundance fit. However, the most recent investigations by
McErlean et al. (\cite{mce98}, \cite{mce99}) and Smith \& Howarth
(\cite{smith99}) indicate that helium enrichment is very likely to be small.

The stellar parameters obtained in this way are given in
Table\,\ref{tab:stellb}.

\subsection{Stellar wind properties}

In the determination of the stellar wind parameters from the Balmer lines we
apply the same strategy as for O-stars (see Puls et al. \cite{pul96}).
We adopt terminal velocities $v_{\infty}$ measured from radiative transfer
fits of the strong UV resonance lines (Howarth et al. \cite{how97},
Haser \cite{has95}, see also Haser et al. \cite{haset} and
\cite{has98}). Then, we calculate unified models varying $\dot{M},
\beta$ and $v_{\rm t}$ to fit the Balmer lines. The strength of the stellar
wind emission over the whole profile is a strong function of $\dot{M}$,
whereas a variation of $\beta$ affects only the central emission core and
its halfwidth. In this way, it is possible to determine both quantities
independently. $v_{\rm t}$ -- as for the analysis of A-supergiants (see
McCarthy et al. \cite{mcc97}) -- influences the relative wavelength
position of the H$_{\alpha}$ emission peak and weakly the width of
H$_{\gamma}$ absorption cores. For weak winds, where H$_{\alpha}$ is purely
in absorption but partially filled in by wind emission, $v_{\rm t}$ has a
crucial influence on the width of the H$_{\alpha}$ absorption core. We note
that a careful adjustment of $v_{\rm t}$ helps to improve the general
appearance of the line profile fit but has only little influence on the
determination of the wind momentum.

\begin{figure*}
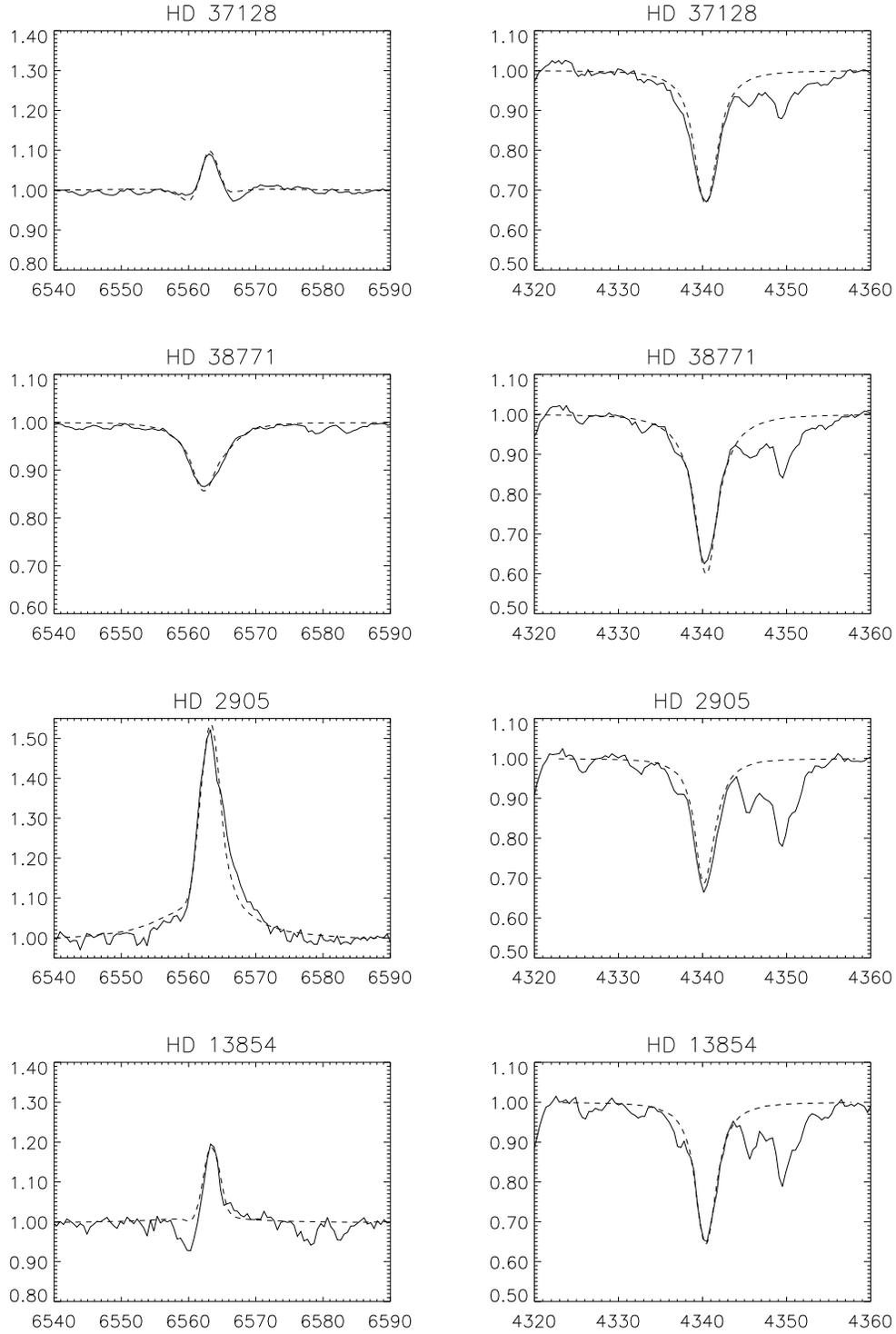

 \centerline{\hbox{
 \psfig{figure=H1519.f1a,height=5cm}
 \psfig{figure=H1519.f1b,height=5cm}
 }}
 \centerline{\hbox{
 \psfig{figure=H1519.f1c,height=5cm}
 \psfig{figure=H1519.f1d,height=5cm}
 }}
  \centerline{\hbox{
 \psfig{figure=H1519.f1e,height=5cm}
 \psfig{figure=H1519.f1f,height=5cm}
 }}
 \centerline{\hbox{
 \psfig{figure=H1519.f1g,height=5cm}
 \psfig{figure=H1519.f1h,height=5cm}
 }}
  \caption{H$_{\alpha}$ (left) and H$_{\gamma}$ (right) line profile fits of 
  the BIa-supergiants of spectral type B0 to B1.
\label{fig:balmb0}}
  \end{figure*}

\begin{figure*}
 \centerline{\hbox{
 \psfig{figure=H1519.f2a,height=5cm}
 \psfig{figure=H1519.f2b,height=5cm}
 }}
 \centerline{\hbox{
 \psfig{figure=H1519.f2c,height=5cm}
 \psfig{figure=H1519.f2d,height=5cm}
 }}

  \caption{H$_{\alpha}$ (left) and H$_{\gamma}$ (right) line profile fits of 
  the B 1.5 Ib supergiants.
\label{fig:balmb1}}
  \end{figure*}

\begin{figure*}
 \centerline{\hbox{
 \psfig{figure=H1519.f3a,height=5cm}
 \psfig{figure=H1519.f3b,height=5cm}
 }}
 \centerline{\hbox{
 \psfig{figure=H1519.f3c,height=5cm}
 \psfig{figure=H1519.f3d,height=5cm}
 }}
  \centerline{\hbox{
 \psfig{figure=H1519.f3e,height=5cm}
 \psfig{figure=H1519.f3f,height=5cm}
 }}

  \caption{H$_{\alpha}$ (left) and H$_{\gamma}$ (right) line profile fits of 
  the B2 Ia-supergiants.
\label{fig:balmb2}}
  \end{figure*}

\begin{figure*}
 \centerline{\hbox{
 \psfig{figure=H1519.f4a,height=5cm}
 \psfig{figure=H1519.f4b,height=5cm}
 }}
 \centerline{\hbox{
 \psfig{figure=H1519.f4c,height=5cm}
 \psfig{figure=H1519.f4d,height=5cm}
 }}

  \caption{H$_{\alpha}$ (left) and H$_{\gamma}$ (right) line profile fits of 
  the B2 Ib supergiants.
\label{fig:balmb3}}
  \end{figure*}

The uncertainties of the stellar wind parameters obtained in this way depend
on the strength of the stellar wind emission. For strong winds leading to a
well pronounced wind emission in H$_{\alpha}$, mass-loss rates can be
determined with an accuracy of twenty percent. In the case of weak winds it
is more difficult to constrain the value of $\beta$, which doubles the
uncertainty of $\dot{M}$. Terminal velocities appear to be uncertain by
fifteen percent. We estimate this number from a comparison of those objects
in common between Howarth et al. (\cite{how97}) and Haser
(\cite{has95}) (note in Table\,\ref{tab:stellb} we use the values given by
Haser \cite{has95}, if available, and Howarth et al. \cite{how97}
otherwise). At the latest spectral type B3 this estimate might be a bit too
optimistic, because of the weakness of wind features in the UV. (For
instance, the $v_{\infty}$ of HD~53138 measured by Haser \cite{has95} is 40
percent smaller than the one determined by Howarth et al.
\cite{how97}). Generally, we expect wind momenta of objects with significant
H$_{\alpha}$ emission to have an uncertainty of 0.1 dex and those of pure
absorption line objects of 0.15 to 0.2 dex.

\begin{figure*}
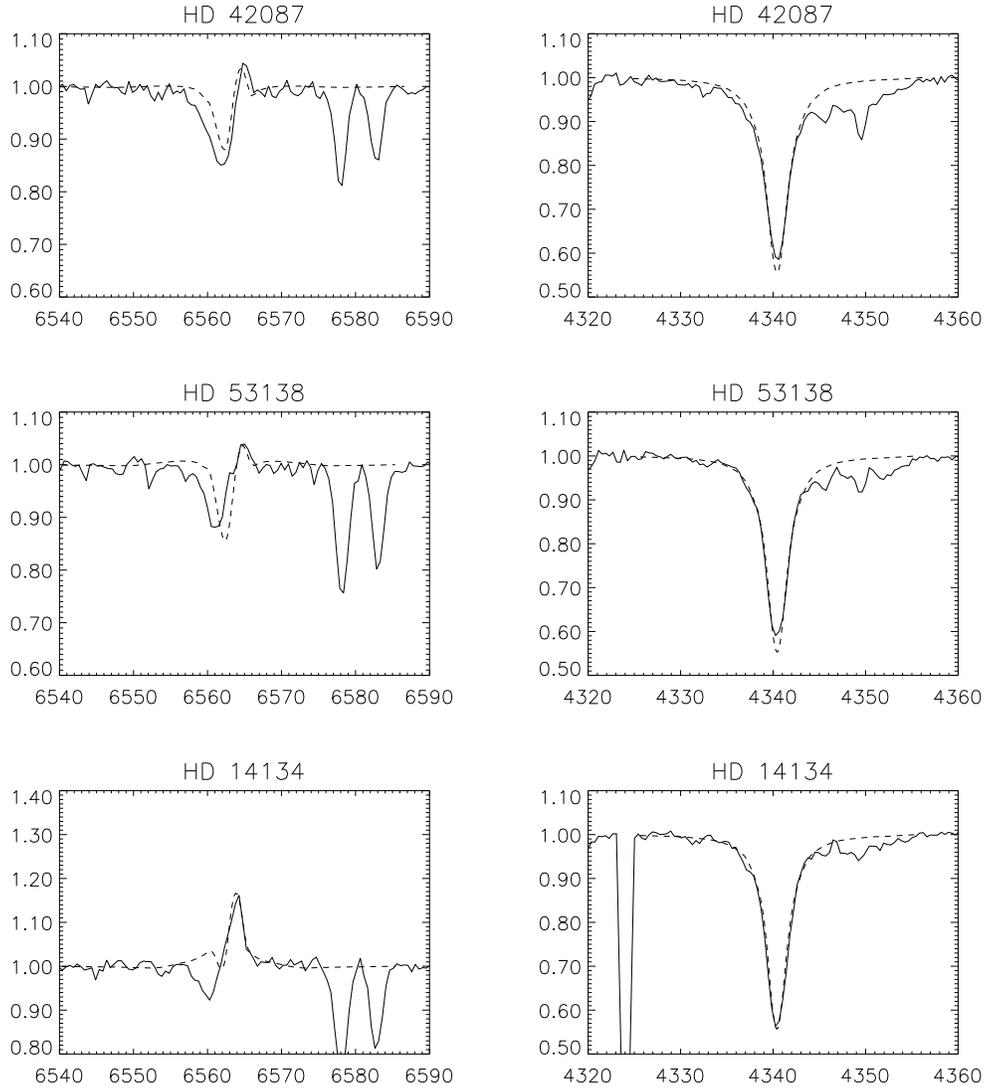

 \centerline{\hbox{
 \psfig{figure=H1519.f5a,height=5cm}
 \psfig{figure=H1519.f5b,height=5cm}
 }}
 \centerline{\hbox{
 \psfig{figure=H1519.f5c,height=5cm}
 \psfig{figure=H1519.f5d,height=5cm}
 }}
  \centerline{\hbox{
 \psfig{figure=H1519.f5e,height=5cm}
 \psfig{figure=H1519.f5f,height=5cm}
 }}

  \caption{H$_{\alpha}$ (left) and H$_{\gamma}$ (right) line profile fits of 
  the B2.5 - and B3 - supergiants.
\label{fig:balmb4}}
  \end{figure*}

The results of the stellar wind analysis are summarized in
Table\,\ref{tab:stellb}. Profile fits of H$_{\alpha}$ and H$_{\gamma}$ are
given in in Figs.~\ref{fig:balmb0},\,\ref{fig:balmb1},\,\ref{fig:balmb2},
\,\ref{fig:balmb3} and \ref{fig:balmb4}. The quality of the H$_{\gamma}$
fits is remarkably good (note that the NLTE profile calculations did not
include blends from the metal lines). In few cases, the values for $v_{\rm
t}$ adopted for H$_{\gamma}$ are smaller than those for H$_{\alpha}$ given
in Table\,\ref{tab:stellb} (HD 193183 and 41117: 10 km/s, HD 206165 and
53138: 15 km/s, HD 14818 and 42087: 20 km/s). As already discussed above, we
attribute this to a stratification of microturbulence through the wind. The
smaller values for H$_{\gamma}$ (as a more photospheric line) are in
agreement with the results obtained by McErlean et al. (\cite{mce99})
in their study of photospheric metal and HeI lines.

In view of the enormous variety of stellar wind strengths encountered, the
H$_{\alpha}$ fits are also satisfactory. However, there are five objects (HD
13854, 41117, 14818, 14143, 14134), where the calculations fail to reproduce
the observed absorption dip blueward of the emission peak. For two objects
(HD 42087, 53138) with slightly weaker winds than those five the absorption
dip is reproduced in its central depth but not in the width towards higher
velocities. This might be the result of the neglect of metal line blanketing
and blocking in our model atmospheres affecting the stratification of the
NLTE H$_{\alpha}$ line source function through the expanding wind. We also
note that we have neglected the influence of a possible He{\sc ii} blend at
these rather low effective temperatures. Another explanation might be
stellar wind variations and deviations from spherical symmetry and
homogeneity leading to temporary absorption dips (see also the discussion in
Puls et al. \cite{pul96}, Kaufer et al. \cite{kauf96}, Kudritzki
\cite{kud99b}).

For three objects of our sample mass-loss rates have been determined from
the continuous free-free stellar wind emission at radio wavelength by
Scuderi et al. \cite{scuderi98}. For HD 37128, HD 2905 and HD 41117
they obtained 2.1, 2.0 and 1.5 $10^{-6}$ M$_{\odot}$/yr, in reasonable
agreement with our H$_{\alpha}$ method. (Note that the distances adopted by
Scuderi et al. are identical to ours, however their value of $v_{\infty}$ 
for HD~37128 is
somewhat higher. Using our value and their observed radio fluxes
we obtain 1.8 $10^{-6}$ M$_{\odot}$/yr for the radio mass-los rate of
HD~37128). For HD~53138 and HD~206165 upper limits of 
4.0 and 6.0 $10^{-7}$ M$_{\odot}$/yr for the mass-loss rate were determined
by Drake \& Linsky \cite{drake89} from radio flux measurements using the VLA.
(We have scaled the Drake \& Linsky  results to our distances and 
values of $v_{\infty}$).
These limits may not be regarded as significant, as they are a factor of 4
and 10 higher than the mass-loss rates from our H$_{\alpha}$ 
determination. However , for HD~53138 Barlow \& Cohen \cite{bar77}
have measured a significant infrared excess at 10 micron requiring a factor 20 
higher mass-loss rate than ours, if caused by free-free wind emission. With
the upper limit from the radio observations we conclude that the IR excess
must have another reason. Indeed,
the inspection of the IRAS Point Source Catalogue for HD~53138 indicates that
the 10 micron flux measured by Barlow \& Cohen is peculiarily high compared
with the values at lower and higher wavelengths. For all other objects in
common with Barlow \& Cohen our mass-loss rates do not lead to severe
discrepancies with the measured IR excesses, if the accuracy of the
photometry is taken into account. (We have also investigated the CDS
Catalog of Infrared Observations, Edition 5 by Gezari et al. and realized that
for many of our targets the IR-photometry shows a sigificant scatter. We
attribute this to the photometric variability of B-supergiants, 
which makes it difficult to measure an IR-excess unless the spectral energy
distribution is observed simultaneouesly.).

\section{Spectral analysis of A-supergiants}

The crucial problem for A-supergiants is to find luminous objects in the
solar neighbourhood with reasonably determined distances. So far, we have
only four objects in our sample and we have to admit that the distance to
the brightest object HD 92207 in Car OB1 is quite uncertain. Originally, it
was assigned a distance modulus of only 11$\fm$1 by Humphreys (\cite{hum70})
although this star was also referenced as a member of the cluster NGC~3324
which is accredited with a distance modulus of 12$\fm$5 by
Lyng\aa~(\cite{lyn70}). Blaha \& Humphreys (\cite{bh89}) obtain a distance
modulus of 12$\fm$0 for Car OB1. The problem with obtaining an accurate
distance to this star is that it lies in the direction of the Car OB1
association and the line of sight is along a galactic spiral arm. The
distance to the association, and to HD~92207 assuming it is a member, has
been discussed quite thoroughly in the literature, most frequently in
connection with the distance to the $\eta$ Car nebula and the very young
clusters Tr14 and Tr16.  Estimates of these distances range from distance
moduli of 12$\fm$3 to 12$\fm$7 (Walborn \cite{wal95}, Massey \& Johnson
\cite{mass93}), while Shobbrook \& Lyng\aa~(\cite{shob94}) summarize the
stellar cluster content and extent of Car OB1 (see their figure 3) showing
that the young star population is distributed in depth over 2 to 3 kpc.  We
therefore adopt a distance modulus of 12$\fm$5 for HD~92207 but acknowledge
that there is a large inherent uncertainty in this value.

Photometric data, cluster membership and spectral type for all objects are
summarized in Table\,\ref{tab:aphoto}. M$_{\rm v}$ values are determined in
the same way as for the B-supergiants.

\begin{table*}
  \caption[]{Spectral types and photometric data of A-supergiants}
  \label{tab:aphoto}
  \begin{center}
    \begin{tabular}{r c c c c c c}
      \hline
 star & spectral type & assoc. & m$_{\rm v}$ & B-V & E(B-V) & M$_{\rm v}$  \\
      \hline
  HD  &               &        &   mag   & mag &  mag   &  mag     \\     
      \hline
  92207 & A0Ia    & Car OB1  & 5.45 &  0.50 & 0.48 & -8.54 \\
  12953 & A1Ia    & Per OB1  & 5.67 &  0.61 & 0.59 & -7.75 \\
  14489 & A2Ia    & Per OB1  & 5.17 &  0.37 & 0.32 & -7.51 \\ 
 223385 & A3Ia    & Cas OB5  & 5.43 &  0.67 & 0.62 & -7.99 \\ 

\end{tabular}
\end{center}
\end{table*}
 
\subsection{Stellar parameters}

As for the B-supergiants, the stellar parameters are obtained from the
quantitative analysis of the stellar spectrum. The model atmosphere fit
using NLTE line formation calculations on top of metal line blanketed Kurucz
atmospheres of the Mg {\sc i/ii} ionization equilibrium and the H$_{\gamma}$
profile yields effective temperature and gravity (see Venn \cite{venn95b},
Przybilla et al. \cite{przy99}). Then, our unified models, which are
unblanketed to reduce the computational effort, are applied to obtain the
stellar wind parameters. To compensate for the neglect of blanketing we have
used effective temperature values 400 K higher than those obtained with
blanketed models in the wind calculations (see also McCarthy et al.
\cite{mcc97}). (Table\,\ref{tab:stella} lists the effective temperatures
obtained from line blanketed models). The radii are then derived exactly in
the same way as for the B-supergiants and also the gravities are iterated
taking into account the influence of winds on H$_{\gamma}$.
(Table\,\ref{tab:stella} contains the $\log g$ values obtained after this
iteration). Normal helium abundance is again adopted for all calculations,
except for HD~92207 for which Przybilla et al. (\cite{przy99}), in
their recent analysis, found an enhanced helium abundance. We adopt
N(He)/N(H)=0.2 for this object.

\begin{table*}
  \caption[]{Stellar parameters and stellar wind properties of A-supergiants}
  \label{tab:stella}
  \begin{center}
    \begin{tabular}{r c c c c c c c c c c}
      \hline
 star & T$_{\rm eff}$ & R$_{\ast}$ & $\log g$ & $\log L/{\rm L}_{\odot}$ & v$_{\rm rot}$ &
v$_{\rm t}$ & $\beta$ & v$_{\infty}$ & $\dot{M}$ & $\log D_{\rm mom}$ \\
      \hline
  HD  &    kK     & R$_{\odot}$&  cgs  &                   &   km/s    &
 km/s   &         & km/s         &$10^{-6} $M$_{\odot}$/yr& cgs \\     
      \hline
  92207 & 9.4 & 192. & 0.95 & 5.41 & 35 & 15 & 1.0 & 235. & 1.31 & 28.43 \\ 
  12953 & 9.1 & 145. & 1.10 & 5.11 & 35 & 10 & 1.5 & 150. & 0.43 & 27.69 \\
  14489 & 9.0 & 128. & 1.45 & 4.99 & 40 & 20 & 2.5 & 190. & 0.14 & 27.27 \\
 223385 & 8.4 & 174. & 0.85 & 5.13 & 30 & 10 & 1.5 & 190. & 0.65 & 27.91 \\
 
\end{tabular}
\end{center}
\end{table*}

\begin{figure*}
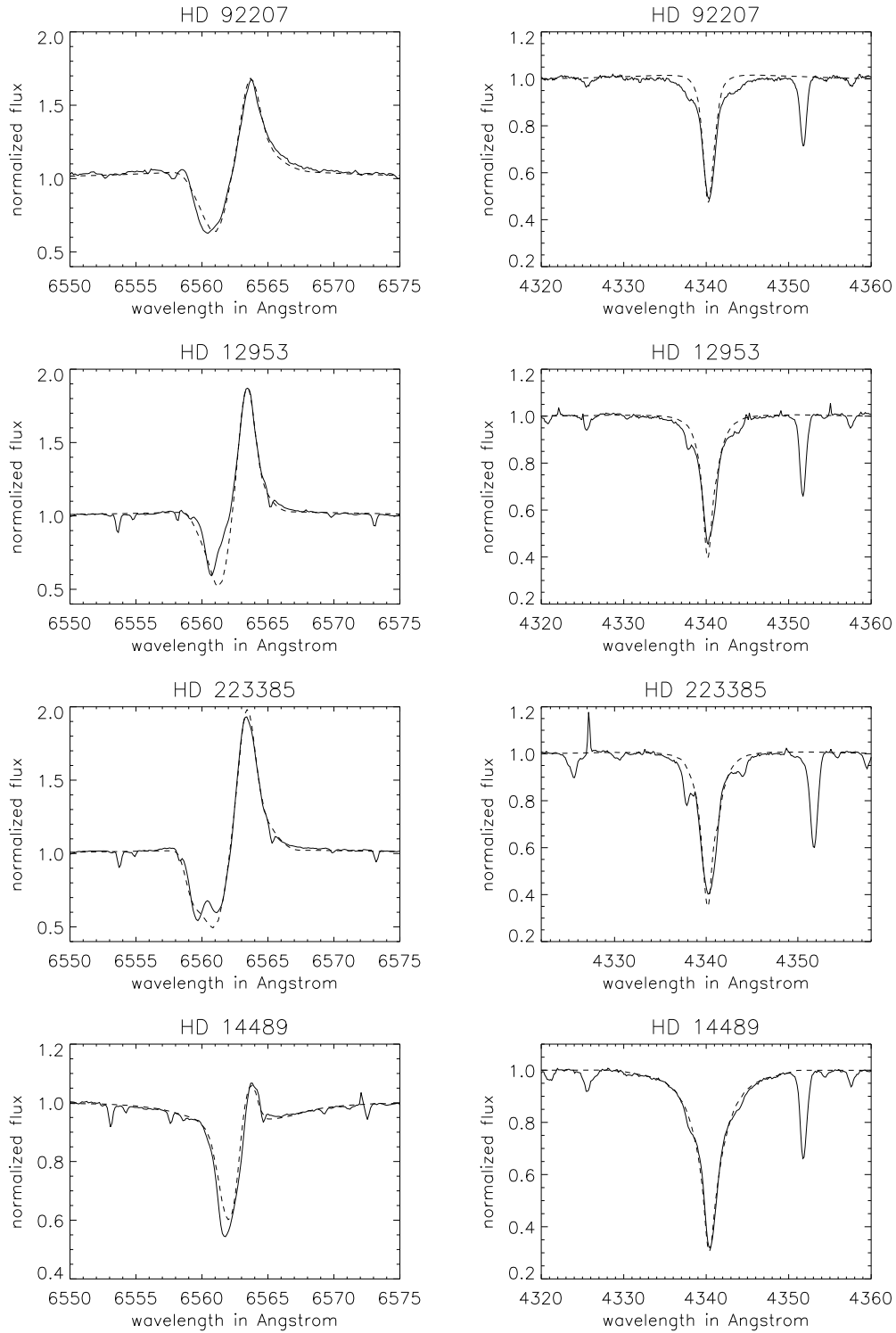

 \centerline{\hbox{
 \psfig{figure=H1519.f6a,height=5cm}
 \psfig{figure=H1519.f6b,height=5cm}
 }}
 \centerline{\hbox{
 \psfig{figure=H1519.f6c,height=5cm}
 \psfig{figure=H1519.f6d,height=5cm}
 }}
 \centerline{\hbox{
 \psfig{figure=H1519.f6e,height=5cm}
 \psfig{figure=H1519.f6f,height=5cm}
 }}
 \centerline{\hbox{
 \psfig{figure=H1519.f6g,height=5cm}
 \psfig{figure=H1519.f6h,height=5cm}
 }}

  \caption{H$_{\alpha}$ (left) and H$_{\gamma}$ (right) line profile fits of 
  the galactic A-supergiants.
\label{fig:balma}}
  \end{figure*}

\subsection{Stellar wind properties}

In the analysis of the Balmer lines to obtain the stellar wind parameters
simultaneously with the gravity we follow exactly the procedure outlined by
McCarthy et al. (\cite{mcc97}). The strength of the H$_{\alpha}$
emission peak determines the mass-loss rate $\dot{M}$, when the width of the
emission peak is used to constrain $\beta$ and the blueshift of the
absorption yields $v_{\infty}$. The gravity follows from the wings of
H$_{\gamma}$. The stellar wind turbulent velocity $v_{\rm t}$ is obtained
from the redshift of the emission peak with regard to the photospheric
radial velocity. As explained in detail by SPH and McCarthy et al.
(\cite{mcc97}) the inclusion of non-coherent electron scattering in the
radiative transfer calculation of the Balmer lines is crucial to obtain a
reliable fit. Fig. \,\ref{fig:balma} gives an impression of the quality of
the fits obtained.

The results of the analysis are compiled in Table\,\ref{tab:stella}. While
the terminal velocities of all four objects are very similar, the mass-loss
rates vary significantly with luminosity. From Fig. \,\ref{fig:balma} one
can learn, how the strengths of the stellar winds influences the Balmer
lines of A-supergiants. The H$_{\alpha}$ profile of HD~14489 -- the object
with the lowest luminosity and largest gravity -- is least affected by
stellar wind emission. The profile fits of the Balmer lines are very
satisfactory. The general P Cygni shape of H$_{\alpha}$ and the photospheric
absorption wings of H$_{\gamma}$ are simultaneously well reproduced. Small
deviations between theory and observation in the line cores are attributed
to stellar wind variability (see Kudritzki \cite{kud99b}) or the
imperfectness of the (still) unblanketed model atmospheres. This will
require further investigations. The electron scattering wings of HD~12953
and HD~223385 are also represented well by the theory. However, the fit of
HD 92207 is somewhat problematic. H$_{\alpha}$ shows very pronounced
electron scattering wings which requires a very low gravity for the fit at
this temperature to have sufficient electron scattering optical depth and
pathlength in the photosphere. However, then these scattering wings are also
weakly present in H$_{\gamma}$ causing a slight deviation from the observed
profile. This may be a problem of the data reduction in terms of continuum
rectification at blue wavelengths but it could also be an indication of the
imperfectness of the present model atmosphere approach. (Since the electron
scattering wings are formed in the hydrostatic photosphere -- see references
mentioned above-- the discrepancy cannot be attributed to a possible
clumpiness in the stellar wind as for objects with very strong mass-loss
such as Wolf-Rayet stars). Enhancing the
gravity by 0.07 dex (see Fig.\,\ref{fig:hd92207}) improves the H$_{\gamma}$
fit significantly but weakens the H$_{\alpha}$ scattering wings. We conclude
that the uncertainties in gravity and mass-loss rate induced by this
disturbing inconsistency are small. However, it will be important to
investigate its physical background in future work.

\begin{figure*}
 \centerline{\hbox{
 \psfig{figure=H1519.f7a,height=5cm}
 \psfig{figure=H1519.f7b,height=5cm}
 }}
  \caption{H$_{\alpha}$ and H$_{\gamma}$ fits of HD 92207 for the same
parameters as in Table\,\ref{tab:stella} but $\log g$ = 1.00 and
$\dot{M} = 1.6$ $ 10^{-6} $M$_{\odot}$/yr. 
\label{fig:hd92207}}
  \end{figure*}

The accuracy of the stellar wind parameters obtained by the H$_{\alpha}$
fits is comparable to McCarthy et al. (\cite{mcc97}). Both mass-loss
rates and terminal velocities can be fitted within 20 percent accuracy
yielding an uncertainty of 0.15 dex for the wind momentum of an individual
object. In addition, the determination of $v_{\infty}$ from H$_{\alpha}$
might lead to results systematically different from the study of UV lines
such as the Mg{\sc ii} resonance lines or Fe{\sc ii} lines arising from the
ground state. So far, there are not enough results published to allow a
careful investigation of this question. However, a comparison with Lamers
et al. (\cite{lam95}), who used IUE high resolution spectra to
determine terminal velocities of blue supergiants, shows acceptable
agreement for the three objects in common, HD~92207 ($200\pm30$ km/s),
HD~12953 ($170\pm20$ km/s) and HD~14489 ($150\pm20$ km/s).

\begin{figure}
 \centerline{\hbox{
 \psfig{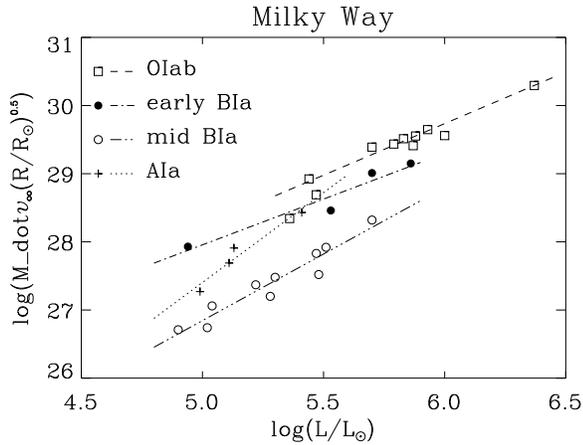}
 }}
  \caption{Wind momenta as function of luminosities for galactic supergiants
  of spectral types O, B (early B: B0 to B1; mid B: B1.5 to B3) and A. The
  straight lines for each group of spectral types are the corresponding
  regression curves.
\label{fig:wlrgal1}}
  \end{figure}

\section{Wind momenta and luminosities}

Modified stellar wind momenta 

\begin{equation}
  D_{\rm mom} = \dot{M}v_{\infty}(R_{\ast}/R_{\odot})^{0.5}
\end{equation}
as function of luminosities are given in Table\,\ref{tab:stellb},
Table\,\ref{tab:stella} and shown in Fig.\,\ref{fig:wlrgal1}. To investigate
the effects of temperature dependence we have included the O Ia and Ib
supergiants from the work by Puls et al. (\cite{pul96}).
Fig.\,\ref{fig:wlrgal1} reveals several trends. Wind momenta are definitely
correlated with the stellar luminosities. However, the wind momentum --
luminosity relationship (WLR) shows a distinct dependence on spectral type.
Wind momenta are largest for the O-supergiants and then decrease with
decreasing effective temperature. The early B spectral types between B0 to
B1 have wind momenta about 0.35 dex smaller than the O-supergiants (note
that the O-supergiant with the smallest wind momentum in
Fig.\,\ref{fig:wlrgal1}, HD~18409, is of spectral type O9.7 Ib. It is not 
included in the calculation of
the O-supergiant regression curves). The spectral types later than B1 (mid B
by our definition) show even weaker winds reduced by roughly 1.1 dex in wind
momentum relative to the O-supergiants. A-supergiants, on the other hand,
have wind momenta between those of early and mid B spectral types.

\begin{table}
  \caption[]{Coefficients of the wind momentum -- luminosity relationship
  for A-, B- and O-supergiants of the solar neighbourhood.}
  \label{tab:wlrgal1}
  \begin{center}
    \begin{tabular}{c c c c}
      \hline
 sp. type & log $D_{0}$ & x & $\alpha_{\rm eff}$ \\

      \hline
    A     & 14.22$\pm$2.41 & 2.64$\pm$0.47 & 0.38$\pm$0.07 \\
   mid B  & 17.07$\pm$1.05 & 1.95$\pm$0.20 & 0.51$\pm$0.05 \\
  early B & 21.24$\pm$1.38 & 1.34$\pm$0.25 & 0.75$\pm$0.15 \\
    O     & 20.40$\pm$0.85 & 1.55$\pm$0.15 & 0.65$\pm$0.06 \\

\end{tabular}
\end{center}
\end{table}

Adopting a WLR of the form

\begin{equation}
 \log D_{\rm mom} = \log D_{0} +  x \log(L/L_{\odot}) ,
\end{equation}
we can determine the coefficients log $D_{0}$ and x by linear regression
(see Fig.\,\ref{fig:wlrgal1} and Table\,\ref{tab:wlrgal1}). The reciprocal
value of the slope x may be interpreted as the effective exponent
$\alpha_{\rm eff}$ describing the depth dependence of the radiative line
force. In the theory of line driven winds $\alpha_{\rm eff}$ can be
expressed as the difference of two dimensionless numbers, $\alpha$ and
$\delta$, the usual force multiplier parameters of the radiative line force.
$\alpha$ corresponds to the power law exponent of the line strengths
distribution function controlling the relative contribution of strong and
weak lines to the radiatiative acceleration, whereas $\delta$ describes the
depth variation induced by ionization changes of the total number of lines
contributing (see Abbott \cite{abb82}, Kudritzki et al. \cite{kud89},
Puls et al. \cite{pul96}, Kudritzki \cite{kud98}, Kudritzki {\it et
al.} \cite{kud98b}, Puls et al. \cite{pul99} for a more detailed
discussion). Typical values for O-stars are $\alpha$ = 0.7 to 0.6 and
$\delta$ = 0.1 to 0.0 .

\begin{equation}
 \alpha_{\rm eff}= 1/x = \alpha - \delta .
\end{equation}
Changes in the slope x of the WLR as function of spectral type are therefore
an additional indication that the winds are driven by different sets of
ions. From Table\,\ref{tab:wlrgal1} (see also Fig.\,\ref{fig:wlrgal1}) we
infer that the WLR is definitely steeper for the spectral types mid B and A
than for O-supergiants (although the value for the A-supergiants is
uncertain because of the small number of objects). This is exactly what one
expects, if spectral lines of the iron group at lower ionization stages
({\sc iii} and {\sc ii}) absorb the photon momentum and drive the wind (Puls
et al. \cite{pul99}). The slope of the early B WLR depends on the
uncertain distance of the low luminosity object HD 38771 and would become
steeper, if the Hipparcos parallax is too large because of the effects of
the Lutz-Kelker bias (see Reid \cite{reid98}). Therefore, the difference to the
O-supergiants in Table\,\ref{tab:wlrgal1} is very likely not significant.

The variation of height $D_{0}$ of the WLR is a more complex matter, since
this quantity depends on the {\it flux weighted} total number of spectral
lines as well as on $\alpha_{\rm eff}$ and $\alpha$ itself (Puls {\it et
al.} \cite{pul99}). Only in cases of similar slopes a comparison of $D_{0}$
gives direct insight into the {\it absolute} number of effectively driving
lines, whereas in all other cases their {\it distribution} with respect to
line-strength ($\alpha$) has a significant influence on the offset.

Due to its (partial) dependence on flux-weighted line number, $D_{0}$ varies
between spectral types not only because of ionization changes but also
because of the different spectral locations of the lines with regard to the
flux maximum and absorption edges such as the hydrogen Lyman - and Balmer -
edges. At least with respect to the observed differences in $D_{0}$ between
O-type and A-type supergiants (accounting for the changes in $\alpha_{\rm eff}$
), this interpretation seems to be correct, as demonstrated by Puls
et al. \cite{pul99}. However, to predict the variation of $D_{0}$
over the whole range of effective temperatures and thus to check whether our
present interpretation is generally valid will require more detailed
calculations of radiation driven wind models in the future. We feel that
especially the pronounced drop in $D_{0}$ between effective temperatures of
23500~K and 22500~K (on our temperature scale of unblanketed models) will
provide a challenge for the theory. A detailed spectroscopic
study of a larger sample of objects in this transition range of
temperatures to disentagle stellar wind properties in more detail will
certainly be extremely valuable. In addition, a careful re-investigation 
whether systematic effects (for instance, deviations of the helium 
abundances from the normal value or metal line blanketing and blocking in
this temperature range) may have influenced the results of the spectroscopic
analysis will be important.

\begin{figure}
 \centerline{\hbox{
 \psfig{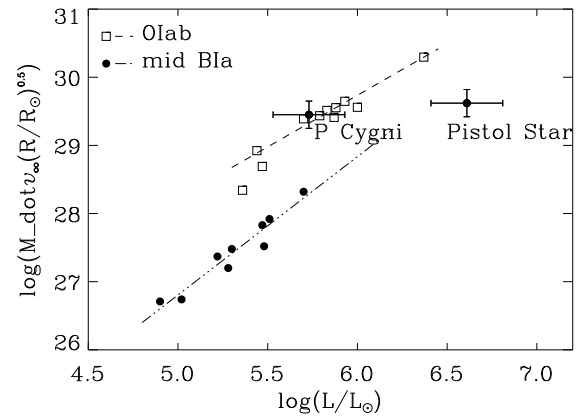}
 }}
  \caption{Wind momentum as function of luminosity for galactic supergiants
  of spectral type O and mid B compared with the wind momentum of the very
  luminous Pistol Star in the Galactic Centre and the extreme Luminous Blue 
  Variable P~Cygni. 
\label{fig:wlrgalpistpcyg}}
  \end{figure}

The observed WLR of ``normal'' supergiants of mid B spectral type allows a
comparison with more extreme objects of comparable spectral type. Very
recently, Figer et al. (\cite{fig99}) have investigated the physical
properties of the spectacular ``Pistol Star'' in the Galactic Centre. In
their detailed spectroscopic study using high quality IR spectra and unified
NLTE model atmospheres with winds they have found that this object has a
luminosity of $\log L/{\rm L}_{\odot}$ = 6.6 and is one of the most luminous
and - therefore - most massive stars known so far. However, despite of the
extreme luminosity the stellar wind of this object is surprisingly weak.
Figer et al. (\cite{fig99}) obtain $\dot{M} = 3.8$ $10^{-5}
$M$_{\odot}$/yr, $v_{\infty} = 95$~km/s and $\log D_{\rm mom} = 29.62$ for
mass-loss rate, terminal velocity and modified wind momentum, respectively.
This is roughly 1.0 dex less than one would expect for an O-star of the same
luminosity. Ignoring the spectral type dependence of the WLR one would be
tempted to conclude that the Pistol Star is probably an unresolved multiple
system of less luminous objects, although the careful investigation by Figer
et al. (\cite{fig99}) does not give any indication in this direction.
However, the effective temperature obtained for this object in their NLTE
analysis is $T_{\rm eff}$ = 14100 K and it is, therefore, more appropriate
to compare with the our new WLR of mid B spectral types. This is done in
Fig.\,\ref{fig:wlrgalpistpcyg}. Given the uncertainties of the complex
analysis of the Galactic Centre IR - data and the fact that the Pistol Star
is somewhat cooler than B3 Ia - supergiants, we conclude from a comparison
of wind momenta that this object may indeed be a single object of extreme
luminosity but with a ``normal'' stellar wind.

On the other hand, we note from Figer et al. (\cite{fig99}) that the
Pistol Star is located in a ``sparsely populated zone in the HRD beyond the
Humphreys-Davidson-limit where unstable stars of the Luminous Blue Variable
type are found''. It is quite surprising that such a star exhibits just a
normal stellar wind. This becomes more evident, if we add the proto-type
Luminous Blue Variable P Cygni to Fig.\,\ref{fig:wlrgalpistpcyg}. This
object has been studied by Lamers et al. (\cite{lam96}) and Najarro
et al. (\cite{naj97}) (see also Pauldrach \& Puls \cite{paul90},
Langer et al. \cite{lan94}) and has an effective temperature of
$T_{\rm eff}$ = 18100 K corresponding to our specral types B2 and B3. With
log $L/{\rm L}_{\odot}$ = 5.75 its luminosity is only slightly higher than
the one of HD 41117 the most luminous ``normal'' mid B-supergiant in our
sample. However, with $\dot{M} = 3.0$ $10^{-5} $M$_{\odot}$/yr, $v_{\infty}
= 185$~km/s and $\log D_{\rm mom} = 29.48$ the wind momentum of P Cygni is
1.2 dex larger than expected from our regression relation for this spectral
type.

The explanation for this behaviour has already been worked out by Lamers
et al. (\cite{lam95}) in their discussion of the ``bi-stability of
B-supergiant winds''. P Cygni is very likely an object that has undergone
substantial mass-loss in its evolutionary history and is therefore rather
close to the Eddington limit (see Langer et al. \cite{lan94}). It is
definitely helium enriched and the rather small terminal velocity leads to
the conclusion that the gravity (and therefore the mass) must be very small.
Langer et al. (\cite{lan94}) estimate $\log g$ = 2.0 from their wind
models which are also able to reproduce the observed high wind momentum. Our
``normal'' objects have significantly higher gravities and are more distant
from the Eddington limit. However, as Pauldrach \& Puls (\cite{paul90}) and
Lamers \& Pauldrach (\cite{lam91}) have shown for the temperature range of
B1.5 to B3 supergiants, the distance to the Eddington limit -- below a
critical threshold -- is crucial for the strengths of stellar winds. Closer
to the Eddinton limit the radiation driven wind becomes slower and denser
which increases the stellar wind optical thickness in the Lyman continuum
and -- in the temperature range between B1.5 and B3 spectral types --
suddenly affects the ionization of iron group elements, which can lead to a
strong increase of stellar wind momentum. We therefore assume that the
difference in gravity (or stellar mass) is the reason for the difference
between the ``normal'' supergiants and the extreme object P Cygni. In this
sense, the pistol star must either be a ``normal'' object sufficiently away
from the Eddington limit or the reduced flux in the Lyman continuum at
$T_{\rm eff}$ = 14000 K (relative to 22000 to 18000 K) leads to a more
moderate stellar wind. Of course, this whole scenario will have to be
confirmed by a comprehensive calculation of stellar wind models in the
appropriate range of stellar parameters.

\section{Stellar wind momenta and extragalactic distances}

\begin{figure}
 \centerline{\hbox{
 \psfig{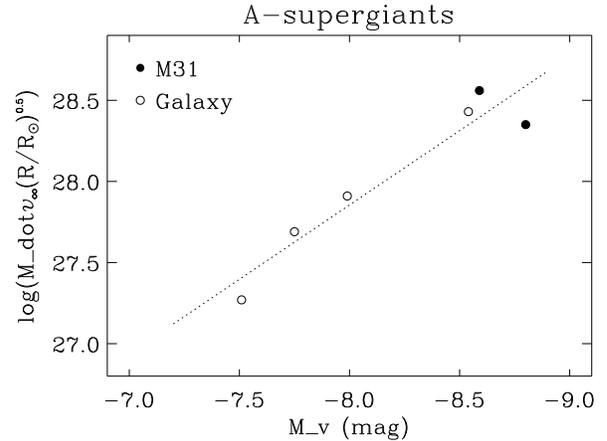}
 }}
  \caption{Wind momentum of galactic and M31 A-supergiants 
  as function of absolute visual magnitude. The data of the M31 objects are
from McCarthy et al. (\cite{mcc97}). The dashed curve is the linear regression obtained from
all objects. 
\label{fig:gaia2}}
  \end{figure}

It has long been a dream of stellar astronomers to use the most luminous
blue stars as individuals to determine the distances to other galaxies.
Evidently, the existence of the WLR provides a unique opportunity to obtain
information about absolute magnitudes from a purely spectroscopic analysis
of the stellar wind lines. To illustrate the situation we show the modified
wind momenta of our A-supergiants as function of absolute visual magnitude
in Fig.\,\ref{fig:gaia2}. We have included the two A-supergiants in M31
studied recently by McCarthy et al. (\cite{mcc97}) in their analysis
of Keck HIRES spectra. The diagram is exciting for several reasons. It
demonstrates that the galactic objects agree well with objects in another
galaxy to form a nice WLR with only a small scatter around the mean
relationship (the fact that the use of absolute magnitude on the abscissa
instead of luminosity leads to a similarily well defined relationship
reflects only that the bolometric corrections are small). In addition, the
relationship holds to absolute magnitudes up to M$_{\rm v} = -9^{\rm m}$,
i.e., objects of tremendous intrinsic brightness easily to detect and to
investigate spectroscopically in distant galaxies.

\begin{figure}
 \centerline{\hbox{
 \psfig{figure=H1519.f11,width=8.7cm}
 }}

  \caption{Residua in distance modulus $\Delta \mu$ obtained for the
  A-supergiants of Fig.\,\ref{fig:gaia2} (see text).
  The dashed curves indicate the standard deviation $\sigma_{\rm A} =
0\fm$30.
 \label{fig:sigmaa}}
  \end{figure}

However, for the use of the WLR as a distance indicator
Fig.\,\ref{fig:gaia2} is somewhat misleading, since the determination of a
mass-loss rate from an H$_{\alpha}$ profile requires already the a priori
assumption about the stellar radius and therefore the distance. Changing the
distance $d$ or the distance modulus $\mu$ of an object in
Fig.\,\ref{fig:gaia2} by

\begin{equation}
   d_{\rm new} = \alpha d_{\rm old}
\end{equation} 
\begin{equation}   
   \Delta \mu = 5 \log \alpha
\end{equation}   
results in in a shift of the ordinate by

\begin{equation}
   \Delta \log D_{\rm mom} = 2 \log \alpha
\end{equation}
since one can show analytically (Puls et al. \cite{pul96}) that the
quantity

\begin{equation}
 Q= \dot{M} (R_{\ast}/R_{\odot})^{-1.5}
\end{equation}
is an invariant of the profile fitting process, i.e. the strength of
H$_{\alpha}$ as a stellar wind line scales exactly with this quantity. For
instance, assuming a larger distance (and therefore radius) for an observed
object one would have to increase $\dot{M}$ so that $Q$ remains constant to
obtain a similar fit of the H$_{\alpha}$ profile.

The above equations allow to use the scatter around the regression curve of
Fig.\,\ref{fig:gaia2} for an estimate of the accuracy in distance modulus
$\mu$ which can be obtained. With the regression fit

\begin{equation}
  \log D_{\rm mom} = a_{\rm A} + b_{\rm A} M_{\rm v}
\end{equation}
($a_{\rm A} = 20.52, b_{\rm A} = -0.916$) we can calculate residuals
$\epsilon_{i}$ from the regression for each individual object $i$

\begin{equation}
  \epsilon_{\rm A}^{i} = \log D_{\rm mom}^{i} - a_{\rm A} -b_{\rm A} M_{\rm v}^{i}
\end{equation}
A change in distance modulus of (see Eq. (7),(8),(11))

\begin{equation}
  \Delta \mu_{i} = 5 \log \alpha_{i} = -\epsilon_{\rm A}^{i}/(0.4 + b_{\rm A})
\end{equation}
would shift each object $i$ exactly on the regression. Hence, these values of
$\Delta \mu_{i}$ can used as an estimate of the accuracy in distance modulus
(see Fig.\,\ref{fig:sigmaa}). The standard deviation obtained from these
residua is $\sigma_{\rm A}$ = 0$\fm$30.

We can apply a similar procedure for the mid B-supergiants. Since bolometric
corrections are larger than for A-supergiants, we use the original WLR with
$\log L/{\rm L}_{\odot}$ on the abscissa and ($a_{\rm B}=17.07$,
$b_{\rm B}=1.95$)

\begin{equation}
   \log D_{\rm mom} = a_{\rm B} + b_{\rm B} \log L/{\rm L}_{\odot}\\
\end{equation}
\begin{equation}
\epsilon_{\rm B}^{i} = \log D_{\rm mom}^{i}-a_{\rm B}-b_{\rm B}\log
L^{i}/{\rm L}_{\odot}\\
\end{equation}
\begin{equation}
\Delta \mu_{i} = 5 \log \alpha_{i} = 2.5\epsilon_{\rm b}^{i}/(b_{\rm B}-1).
\end{equation}
The corresponding residua $\Delta \mu_{i}$ are shown in
Fig.\,\ref{fig:sigmab}. The standard deviation is $\sigma_{\rm B}$ =
0$\fm$37.

This confirms the estimate by McCarthy et al. (\cite{mcc97}) that with
ten to twenty objects per galaxy, distance moduli as accurate as 0$\fm$1
should be achievable. The idea is to carry out multi-object spectroscopy
with 8m-class telescopes combined with HST photometry to obtain stellar
parameters ($T_{\rm eff}$, $\log g$, abundances) and reddening in a first
step. Then determination of $Q$ from H$_{\alpha}$ (for A-supergiants one
would also obtain $v_{\infty}$ from H$_{\alpha}$) and the application of a
properly calibrated WLR would yield the absolute magnitude. Comparison with
the de-reddened apparent magnitude would, finally, allow to determine the
distance.

The uncertainties in WLR distance moduli appear to be comparable to
those obtainable from Cepheids in galaxies. The advantage of the WLR-method,
however, is that individual reddening (and therefore extinction) as well as
metallicity (see discussion below) can be derived directly from the spectrum
of every object. Moreover, it is a new independent primary method for
distance determination and can contribute to the investigation of systematic
errors of extragalactic distances. However, the crucial question is out to
which distances will the method be applicable.

\begin{figure}
 \centerline{\hbox{
 \psfig{figure=H1519.f12,width=8.7cm}
 }}

  \caption{Residua in distance modulus $\Delta \mu$ obtained for the mid 
  B-supergiants of Fig.\,\ref{fig:wlrgal1} (see text).
  The dashed curves indicate the standard deviation $\sigma_{\rm B}$ =
0$\fm$37.
 \label{fig:sigmab}}
  \end{figure}

The best spectroscopic targets at large distances are certainly
A-supergiants such as shown in Fig.\,\ref{fig:gaia2}. Since massive stars
evolve at almost constant luminosity towards the red, A-supergiants are the
optically brightest ``normal'' stellar objects because of the effects of
Wien's law on the bolometric correction. In addition, we can determine the
wind momenta of these objects solely by optical spectroscopy at H$_{\alpha}$
(without need of the UV). This means that we can use ground-based telescopes
of the 8m class for spectroscopy rather than the ``tiny'' HST (which is then
needed for accurate photometry only).

From Fig.\,\ref{fig:gaia2} we see that the brightest A-supergiants have
absolute magnitudes between -9$^{\rm m}$ and -8$^{\rm m}$. Such objects
would be of apparent magnitude between $20^{\rm m}$ and $21^{\rm m}$ in
galaxies $6\,$Mpc away, certainly not a problem for medium ($2\,$\AA )
resolution spectroscopy with $8\,$m class telescopes. Even in a galaxy like
M100 at a distance of $16\,$Mpc (Freedman et al. \cite{fre94b};
Ferrarese et al. \cite{fer96}) these objects would still be accessible
at magnitudes around $22\fm5$ and would yield wind momentum distances.
Indeed, the HST colour magnitude diagram published by Freedman et al.
(\cite{fre94a}) may show the presence of such objects in M100.

One might ask, of course, whether a medium resolution of $2\,$\AA\ would
still allow a sufficiently accurate determination of effective temperature,
surface gravity, abundances, and wind momentum. We note that the rotational
velocities of A-supergiants are typically on the order of $40\,$km\thinspace
s$^{-1}$, a broadening which is matched by a spectral resolution of roughly
$1.4\,$\AA\ at H$_{\alpha} $. Therefore we expect that the medium resolution
situation will not be dramatically worse, provided sufficient S/N can be
achieved and accurate sky-, galaxy-, and H{\sc ii} region background
subtraction can be performed at very faint magnitudes. Experiments with
observed spectra degraded with regard to S/N, resolution and sky emission
indicate that such observations are challenging but feasible.

\section{Future work}

In summary of the previous sections it is certainly fair to conclude that
the concept of the spectral dependend WLR appears to be an excellent tool to
discuss the strengths of stellar winds and to use the most luminous
``normal'' blue supergiants as extragalactic distance indicators. However,
it became also clear that what has been presented here can only be regarded
as a first step. Investigations in many directions will be neccessary to
establish the WLR-concept reliably. In the following we discuss the most
important next steps.

\paragraph{Galactic calibration.}
So far the galactic WLR-calibration is based only on a small number of objects
with reliable distances per spectral type group. The numbers are
particularily small for early B-supergiants and A-supergiants. In the near
future before the next astrometric space projects become reality, the only
possibility to improve the situation is the careful re-investigation of
existing catalogues with regard to cluster and association membership and
improved reddening corrections. In addition, one might use the progress
obtained with alternative distance indicators such as eclipsing binaries
(Clausen \cite{clau99}), Cepheids using the IR-Baade-Wesselink-Method
(Gieren \cite{gier99}) etc. to use Local Group galaxies like M31 and M33 as
additional calibrators.

\paragraph{Spectral Variability.}
It is well known that H$_{\alpha}$-profiles of A- and B-supergiants vary as
a function of time (see, for instance, Kaufer et al. \cite{kauf96}).
This does, of course, mean that wind momenta derived in the way as described
above will also vary and, therefore, affect the WLR. Very recently,
Kudritzki (\cite{kud99b}) has discussed the problem for the case of the
A-supergiant HD 92207 and found that the uncertainty in wind momentum
introduced by the spectral variability of this object is smaller than 0.15
dex. He concluded that part of the intrinsic scatter of the WLR might be
caused by variability but that the concept of the WLR is not affected by
uncertainties of this magnitude, in particular, if many objects are
investigated per galaxy. However, so far this conclusion is based on the
investigation of one object only. Many others studies will have to be
carried out as function of spectral type and luminosity class to confirm or
to disprove this conclusion.

\paragraph{Metallicity Dependence.}
Since the winds of blue supergiants are driven by absorption of photon
momentum through metal lines, stellar wind momenta must depend on
metallicity in a sense that wind momenta become weaker with decreasing
metallicity. For normal O-stars, this metallicity dependence has been
discussed carefully by Abbott (\cite{abb82}), Kudritzki et al.
\cite{kud87}, Puls et al. (\cite{pul96} and \cite{pul99}).
Observations of O-stars in the Magellanic Clouds with HST have led to a
general confirmation of the theory (see Puls et al. \cite{pul96}).
However, for supergiants of spectral type A and mid B only little work has
been done so far. Observing two A-supergiants in M33 with Keck and HIRES and
making use of the strong metallicity gradient in this galaxy, McCarthy {\it
et al.} (\cite{mcc95}) were able to demonstrate that drastic metallicity
effects are encountered, if the metallicity drops below the SMC value. In
addition, a few B-supergiants have been studied in the Magellanic Clouds
(Kudritzki \cite{kud98}). However, a systematic survey in the Magellanic
Clouds, NGC 6822, M31 and M33 is now under way by the authors of this paper
and their collaborators to close this gap and to provide an empirical
calibration of WLR metallicity dependence. In addition, theoretical work
will be carried out for these spectral types as well.

\paragraph{Wind momenta beyond the Local Group.}
This is the final goal of the project. As discussed above, quantitative
spectroscopy of blue supergiants beyond the Local Group to determine the
strengths of stellar winds and wind momenta is feasible with 8m-class
telescopes and medium resolution spectrographs. Whether or not this will
lead to a new, independent determination of extragalactic distances is open
until the prove is made by the appropriate analysis of the first set of
spectroscopic multi-object observations in a galaxy clearly beyond the Local
Group. These observations will become available soon.

\begin{acknowledgements}

RPK likes to thank the director and staff of Steward Observatory, Tucson,
for their hospitality and support during a sabbatical where much of the
stellar wind analysis has been carried out. He also wishes to thank Norbert
Przybilla and Oliver Kn\"orndel for discussion and support. The funding 
through the ``Verbundforschung Astronomie'' by the Bundesminister f\"ur 
Forschung und Bildung and the DLR under grant 50 R93040 is gratefully 
acknowledged (DJL, JR). The constructive remarks of the referee, Dr. Paul
Crowther, have helped to improve the paper and are very much acknowledged.

\end{acknowledgements}

\end{document}